\documentstyle[aps,prl]{revtex}

\input{epsf}

\begin{document}
 
\twocolumn[\hsize\textwidth\columnwidth\hsize\csname
@twocolumnfalse\endcsname

\title{
	Timing, ``Accidentals'' and Other Artifacts in EPR
	Experiments\cite{PRL/submission}}
\author{Caroline H Thompson\cite{CHT/email}\\
	Dept. of Computer Science, Univ. of Wales, 	Aberystwyth, 
	SY23 3DB, U.K.}

\date{\today}
\maketitle

\begin{abstract}
Subtraction of ``accidentals'' in Einstein-Podolsky-Rosen experiments frequently changes results compatible with local realism into ones that appear to demonstrate non-locality. The validity of the procedure depends on the unproven assumption of the independence of emission events.  Other possible sources of bias include enhancement, imperfect synchronisation, over-reliance on rotational invariance, and the well-known detection loophole.  Investigation of existing results may be more fruitful than attempts at loophole-free Bell tests, improving our understanding of light.
\end{abstract}

\vskip2pc]

Real EPR experiments are very different both from Einstein, Podolsky and Rosen's original idea \cite{Einstein/Podolsky/Rosen35} and from Bell's idealised situation \cite{Bell64}. The general scheme, which can have either one or two detection channels on each side, is described in a multitude of papers \cite{Clauser/Shimony78,Selleri88}.  Bell assumed initially only schemes in which detection in one or other channel of a two-channel experiment was certain.  He also assumed that the ``particles'' came in pairs that could be identified unambiguously.  Others later attempted to modify his ideas to cover feasible experiments, which to date have all (with no credible exceptions) involved light.  The
modifications have been couched in terms of a particle model of light.  A definitive paper on the subject is that of Clauser and Horne \cite{Clauser/Horne74}.  It is found that in all actual experiments auxiliary assumptions have to be made.  The ones that have received most attention are those of ``fair sampling'' and ``no enhancement''.  That these may not be valid is well known \cite{Santos85+,Pascazio89,Thompson96}. This paper is primarily concerned with lesser-known assumptions that receive little or no mention in published papers. I illustrate them drawing on material from Freedman's and Aspect's PhD theses \cite{Freedman72,Aspect83-prl}. Though some of my points are specific to their atomic cascade sources, most apply equally to recent experiments using parametric down conversion.

\begin{table}
\label{table1}
        \begin{tabular}{cccc}
         & Test Statistic  & Upper Limit & Auxiliary\\
        & & & Assumption\\
\hline
Standard & $S_{Std} = 4(\frac{x - y}{x + y})$ & 2 & Fair sampling\\
Visibility & $ S_V = \frac{max + min}{max - min} $ & 1.71 & "\\
CHSH & $S_{C} = 3\frac{x}{Z} - \frac{y}{Z} - 2\frac{z}{Z}$ & 0 &
        No enhancement\\
Freedman & $S_F = \frac{x - y}{Z}$ & 0.25 &  "\\
\end{tabular}
\caption{Various Bell inequalities, for rotationally invariant,
factorisable
experiments.  $x = R(\pi/8)$, $y = R(3\pi/8)$,
$z = R(a, \infty)$ and $ Z = R(\infty,
\infty)$, using the usual terminology
in which $R$ is coincidence rate, $a$ is polariser setting, and
$\infty$ stands for absence of polariser.}
\end{table}
I give in table I formulae for the Bell tests that are commonly used.
More general ones are needed if the source is not rotationally
invariant.

The standard inequality, used for two-channel experiments, is covered in an earlier paper, ``The Chaotic Ball: an Intuitive Analogy for EPR Experiments'' \cite{Thompson96}.  Realist models that infringe it are easily constructed if, as I consider must always be the case, there are ``variable detection probabilities''.  Tests of visibility employ the same assumptions as the rotationally invariant form of the standard test, so they too are invalidated if detection probabilities can vary with ``hidden variable''.  The current paper presents ideas on factors that can explain violations of the final two tests.  Of course, all the factors can play contributory roles in any experiment, regardless of the test actually used.

Single-channel experiments, involving the CHSH or Freedman tests \cite{Aspect81-2}, differ from the others in that the ``detection loophole'' cannot on its own cause violations of the inequalities.  What is needed is failure of one of the other assumptions, or, as it emerges, over-reliance on theories that suggest that emission events are independent.  The tests as given in Table I all rely on rotational invariance, and I am currently looking into some cases in which this may have been assumed on insufficient evidence.  As mentioned earlier, the assumption of ``no enhancement'' may fail.  It is also possible to have imperfect factorability, if there are synchronisation problems \cite{Fine82+}.   The possibility of ``coherent noise'' causing spurious increases in correlations is amongst those considered by Gilbert and Sulcs \cite{Gilbert/Sulcs96}.

But there is another factor that appears to have been almost totally ignored, except within PhD theses.  It is not strictly speaking anything to do with Bell tests, but is a matter of experimental procedure.  Marshall, Santos and Selleri challenged Aspect's logic in subtracting ``accidentals'' before analysing coincidences.  Aspect and Grangier responded with a paper \cite{Aspect/Grangier85} that gave theoretical arguments supporting the practice, and also quoted figures from one of Aspect's experiments that violated an inequality even without the subtraction.  I consider that the matter should have been taken further.  Instead of theoretical arguments, more experimentation is needed.  The figures that Aspect gave were from his {\em two-channel} experiment \cite{Aspect82-0}, that used the standard test, easily violated if detection rates were not constant.

To put the subtraction issue in perspective, let us look its effect on some Bell tests derived from data from Aspect's first EPR experiment (table II).  The raw figures do not infringe any tests: the ``corrected'' ones do.

Detailed figures for his other single-channel experiment are not available, but from the information that is given it seems unlikely that a Bell test would have been violated had ``accidentals" not been subtracted.  It is by no means only Aspect's results that are affected: it has become almost standard practice to do the subtraction, which has similarly important consequences in, for example, the recent Geneva experiments that showed high correlations over a distance of 10 kilometers \cite{Tittel97}.

The mechanism whereby accidentals can cause violation is very straightforward. We can be fairly confident that Bell's inequalities will hold for the raw data, but whether or not they hold after subtraction depends on whether or not true and accidental signals are {\em independent}.  This depends on factors such as correlations between neighbouring emissions, how the detector responds to close or overlapping signals, and instrument dead times.  The number of accidentals, as measured by the number of coincidences when one stream is delayed by, say, 100 ns, is proportional to the product of the numbers of signals on each side.  If the detectors are ``correctly'' adjusted, so that they register half the number of hits when a polariser is inserted (which follows from Malus' Law {\em provided noise and settings of various voltages are appropriate} \cite{Malus_Law}), it is easily seen that the 1:2:4 proportions seen in Table II are just as expected, and that subtraction will always increase all test statistics and hence the likelihood of violations.

\begin{table}
\label{table2}
\begin{tabular}{cccccccc}

        & $x$ & $y$ & $z$ & $Z$ & $S_{Std}$ & $S_C$ & $S_F$\\
\hline
     Raw coincidences & 86.8 & 38.3 & 126.0 & 248.2 & 1.55 & -0.121 &
0.195 \\
        Accidentals & 22.8 & 22.5 & 45.5 & 90.0 &  &  &  \\
        ``Corrected" & 64.0 & 15.8 & 80.5 & 158.2 & 2.42 & 0.096 & 0.309
\\
\end{tabular}
\caption{Effect of standard adjustment for accidental coincidences.
Derived from table VII-A-1 of Aspect's thesis.}
\end{table}

Let us review the whole question of {\em time} in EPR experiments.  The question of accidentals is inextricably entangled with matters of timing, and, besides, this is of interest in its own right.

\begin{figure}
        \centering
        \leavevmode
        \epsfbox{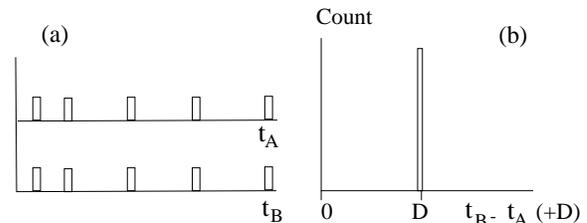}
        \caption{Assumed ideal timing and resultant spectrum.  $D$ is
the delay applied to the $B$ channel so as to place the peak in the centre of the picture.}
        \label{fig1}
\end{figure}
  
Proofs of Bell inequalities do not mention time.  They assume that the source produces pairs of particles and that identification of the pairs poses no problem.  Thus the stream of signals arriving at the coincidence monitor can be envisaged as in Fig.~\ref{fig1}, which also shows the expected time-spectrum --- a single bar whose height is the number of coincidences.  Even this simple picture might have a slight complication, if the pairs are supposed to be produced at completely random times.  There is then a very slight possibility of a second $B$ arrival at {\em any} interval after the $A$ (including in the same ``time-bin''), so that, if we only in fact detect a small fraction of the signals, we expect a low, almost constant, background of  ``accidentals''.  If, however, the physics is such that we {\em
never} get two emissions very close together, then, whatever the spectrum may look like, there can be {\em no} accidentals contributing to the peak unless time-bins are larger than the minimum separation.

\begin{figure}
        \centering
        \leavevmode
        \epsfbox{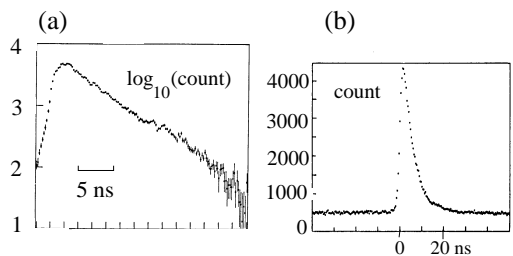}
        \caption{Actual time spectra.  (a) Freedman, (b) Aspect.
Freedman's  could have been slightly distorted by the instrumentation.  It is drawn by hand and represents the combined results of a whole series of runs.  Aspect's is for just one run and would have been displayed on a VDU (most spectra had greater scatter, being collected over shorter periods)}
        \label{fig2}
\end{figure}

Consider now some spectra from actual experiments, Freedman's of 1972 \cite{Freedman72} and Aspect's of around 1980 (Fig.~\ref{fig2}).  Numbers of coincidences can be estimated by defining an ``integration window'' and organising electronics so as to count all signals that arrive within it.  Aspect in his final experiment used a window from $-3$~ns to $+17$~ns.  Freedman used one of just 8~ns length, but does not tell us how the start was chosen.  

From their PhD theses, it is clear how the experimenters were modelling their time-spectra.  Both saw the decreasing region as showing the distribution of emission times of the second ``photon", this being regarded as a particle.  They took the rising front as due to {\em random} ``time jitter''.  There was a constant underlying background of accidentals.

Under this model, there is no possibility of a correlation between detection time and polariser setting, so that the choice of window size cannot in itself cause bias.  It was quite legitimate for the experimenters to chose parameters with a view simply to minimising the running time to achieve a given accuracy.  If, as Aspect thought, the window included 97\% of the true coincidences, there was unlikely in any case to be a problem.    But in truth there may be problems, and in Freedman's case, with a very small window, they may have been sufficiently serious to account for the observed violation of his inequality.  And it is possible that Aspect's estimate of the percentage included was wrong: it may have been based on just {\em one} particular spectrum.  His model did not suggest to him that they might have varied in shape.  He could not, without due motivation, have judged the shape just by eye, owing to the ``accidentals'' and to the high scatter of the spectra in his production runs.

I have identified three types of potential problem.  Firstly, there can be a small time difference when we add or remove a polariser.  Aspect would have been able to correct for this using a variable time delay, but it is possible that Freedman did not.  Secondly, the assumption of random time jitter, accepted within quantum theory (QT) from the 1920's \cite{Lawrence/Beams28}
could be wrong.  Thirdly, there is a factor, ruled out by Aspect and Freedman's model, that can potentially produce errors of the same order of magnitude as the known time jitter (standard deviation around 1 ns) but with a definite correlation with polariser setting.  It depends on a pure wave model of light.

For under a purely wave view of light, the most natural interpretation of the observed spectra is that the $A$ and $B$ ``photons" are emitted {\em simultaneously}, and each is a wave that starts at high intensity and decreases at roughly negative exponential rate, only the $A$ one much faster than the $B$ \cite{Detection}. There is possible experimental support for this view: it implies that there would be the possibility of multiple detections of a single ``photon'' if the electronics did not restrict us to the {\em first} detection only.  This might explain Aspect's problems with ``{\em post-impulsions}'', some of which occurred despite dead times of 16~ns or more.

In this wave model, polarisers have the effect of reducing the intensity of each individual signal.  If the actual process of detection requires only a very short signal, then each complete ``photon'' has many chances of detection.  If it has passed through a polariser, therefore, the probability at each possible time will be reduced, resulting in the time of detection tending to be later.  This can affect the shape of time spectra and, unless very large windows are used, the logic of EPR experiments.  For when polarisers are parallel, we shall tend to get positive correlations between $A$ and $B$ intensities, translating into positive correlations in detection times and good synchronisation.  When polarisers are orthogonal, synchronisation will be relatively poor and time differences that are too large to be recognised as coincidences more common.  This will have the net effect of increasing the visibility of the coincidence curve.  In the accepted language of EPR experiments,  we do not have exact factorability \cite{Fine82+}.

That experimenters do not recognise this possibility is evident from the fact that they consider two quite distinct methods of estimating the ``lifetime of the intermediate state of the cascade'' to be equivalent.  It can, it is thought, be estimated equally reliably by measuring the slope of the spectrum obtained {\em either} as above {\em or} (as given in a reference from Aspect's thesis) by a method, in which many ``photons'' are detected simultaneously and produce directly a time-varying electric signal \cite{Havey77}.  Further experimentation in this area might be very rewarding.

From the point of view of EPR experiments, though, timing is unlikely to be as important as other factors.  It is important mainly indirectly, in that the spread of the signal in time obscures information about the intervals between signals and makes the assessment of accidentals ambiguous.

\begin{figure}
        \centering
        \leavevmode
        \epsfbox{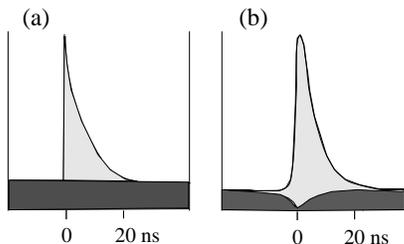}
        \caption{Models of time spectra: (a) quantum mechanics
assumption
and (b) conjectural realistic model.  Light shading: ``true''; Dark
shading: ``accidental'' coincidences.}
        \label{fig3}
\end{figure}

Let us return now to the matter of ``accidentals''.  Aspect's experiments involved large numbers of them, if we take as definition the coincidences obtained when we apply a delay to one channel.  As he says, there could typically be 600 to every 200 ``true'' coincidences displayed on the VDU.  One can  question whether it is possible to extract a valid Bell-type test from the data. His idealisation is illustrated by Fig.~\ref{fig3}(a), taken from his thesis.  But we have no independent way of judging the true picture.   This might, if it exists at all, be as in Fig.~\ref{fig3}(b) --- an idea that cannot be dismissed as entirely {\em ad hoc}: the zero position {\em is} logically different from the remainder.  We cannot say without further justification that the source is not behaving more like, say, a wind instrument, in which only one main note is generally produced at a time (or, in this case, exactly {\em two} notes at a time).  We have stimulating lasers with high coherence properties illuminating a small source containing ionised calcium atoms.  If QT is correct, these atoms are acting independently and subtraction of accidentals is justifiable.  Under QT, then, it
should be possible to continue to violate Bell inequalities when intensities are reduced, so that emissions become well spaced and accidentals negligible.  An experimental test of this is urgently needed, both for purposes of Bell tests and for a better understanding of light.

So why have EPR experiments been so widely accepted as supporting QT, when there are perfectly straightforward local realist possibilities, just a few of which I have introduced in this paper?  It is evident that conventional classical explanations are wrong, as they give wrong predictions.  With the known ``conceptual difficulties'' of QT --- the apparent non-locality --- it seems evident that {\em all} existing theory needs to be challenged.  Aspect, in his thesis, said  that agreement with QT was a privileged method for confirming that the apparatus was correctly set.  Freedman concluded his thesis with a remark to the effect that there was no need to search too hard for causes of systematic error as Bell's inequalities had been violated and were of such general applicability.  Many workers have allowed themselves to be influenced by the fact that various imperfections bring QT predictions nearer to classical ones, reducing the visibility of coincidence curves.  This is true of {\em QT predictions}.  What happens in reality, though, is rather the reverse.  The classical ideas presented in this paper show that there are certain imperfections that {\em increase} visibilities.  If we do not believe in magic, then we must recognise that the experimenters, apart from the very few exceptions such as Holt and Pipkin \cite{Holt/Pipkin74}, have been deceiving themselves.

The explanation for this whole phenomenon lies in sociological and psychological factors --- confusion caused by working with a counterintuitive theory, the pressure to produce results acceptable to peers, the conviction that nobody else has yet found fault with QT.  One ``success'' in EPR experiments has led to another, but the faults have been propagated instead of weeded out.  

Yet we can rescue some very positive results from this story.  What have we actually found?  That we cannot design loophole-free Bell tests using {\em light} --- we have been attempting the impossible.  Why should this be?  If we analyse the experiments carefully, we find that it is because the whole enterprise was undertaken on a false premise: that light could be modelled as {\em particles}.  This is one message.  I believe we have also learned another: that we cannot demonstrate ``quantum entanglement'' by macroscopic experiments.  This phenomenon remains an uncorroborated prediction of QT.

\bibliographystyle{prsty}

\begin{references}
\bibitem{Malus_Law}
Malus' Law should not be assumed in EPR experiments.  It may well be that it holds exactly for {\em electromagnetic intensities}, but a realist theory of detection must admit the likelihood that the probability of detection is {\em not} exactly proportional to the intensity.  Hence Malus' Law would not in general hold exactly for {\em counts}.
\end{references}

 
\end{document}